\documentclass{PoS}

\newcommand{\be}{\begin{eqnarray}}
\newcommand{\ee}{\end{eqnarray}}
\newcommand{\ba}{\begin{array}}
\newcommand{\ea}{\end{array}}
\newcommand{\bi}{\begin{itemize}}
\newcommand{\ei}{\end{itemize}}

\title{Exclusive production of a large mass photon pair  }

\ShortTitle{Exclusive production of a   photon pair  }

\author{A.~Pedrak \footnote{speaker}\\
 National Center for Nuclear Research (NCBJ), 02093 Warsaw, Poland}

\author{B.~Pire \\
          CPHT,  CNRS, \'{E}cole Polytechnique, I.P. Paris,  91128 Palaiseau, France
}

\author{L.~Szymanowski \\
          National Centre for Nuclear Research (NCBJ), 02093 Warsaw, Poland
}

\author{ J.~Wagner \\  National Center for Nuclear Research (NCBJ), 02093 Warsaw, Poland}

\abstract{The scattering amplitude for photoproduction of a large invariant mass diphoton in the generalized Bjorken regime has a very peculiar and interesting analytical structure. The leading twist leading order amplitude is proportional to valence quark  generalized parton distributions taken at the border value $x=\pm \xi$. Cross section estimates show that this process is measurable at JLab energies.  The angular asymmetry triggered by a linearly polarized photon beam is large.   

}

\FullConference{XXVII. International Workshop on Deep-Inelastic Scattering and Related Subjects,\\
		 April 8-12, 2019\\
		Torino, Italy}

\begin{document}

\section{Introduction}

The exclusive photoproduction of two photons on a   unpolarized proton or neutron target
\begin{equation}
\gamma(q,\epsilon) + N(p_1,s_1) \rightarrow \gamma(k_1,\epsilon_1) +  \gamma(k_2,\epsilon_2)+ N'(p_2,s_2)\,,
\label{process}
\end{equation}
 in the kinematical regime of large invariant diphoton mass  $M_{\gamma\gamma}$ of the final photon pair and small momentum transfer $t =(p_2-p_1)^2$ between the initial and the final nucleons, has a number of interesting features \cite{Pedrak:2017cpp}.  First, it is a purely electromagnetic process at Born order - as are deep inelastic scattering (DIS), deeply virtual Compton scattering (DVCS) and timelike Compton scattering (TCS) - and, although there is no deep understanding of this fact, this property is usually accompanied by early scaling. Second, the process is insensitive to gluon GPDs and to singlet  quark GPDs because of the charge symmetry of the two photon final state and is thus very complementary to DVCS and TCS. We thus believe that this reaction may help us to progress in the understanding of hard exclusive scattering in the framework of the QCD collinear factorization  of hard amplitudes  in terms of generalized parton distributions (GPDs) and hard perturbatively calculable coefficient functions

\section{The scattering amplitude}
\begin{figure}
\includegraphics[width=8cm]{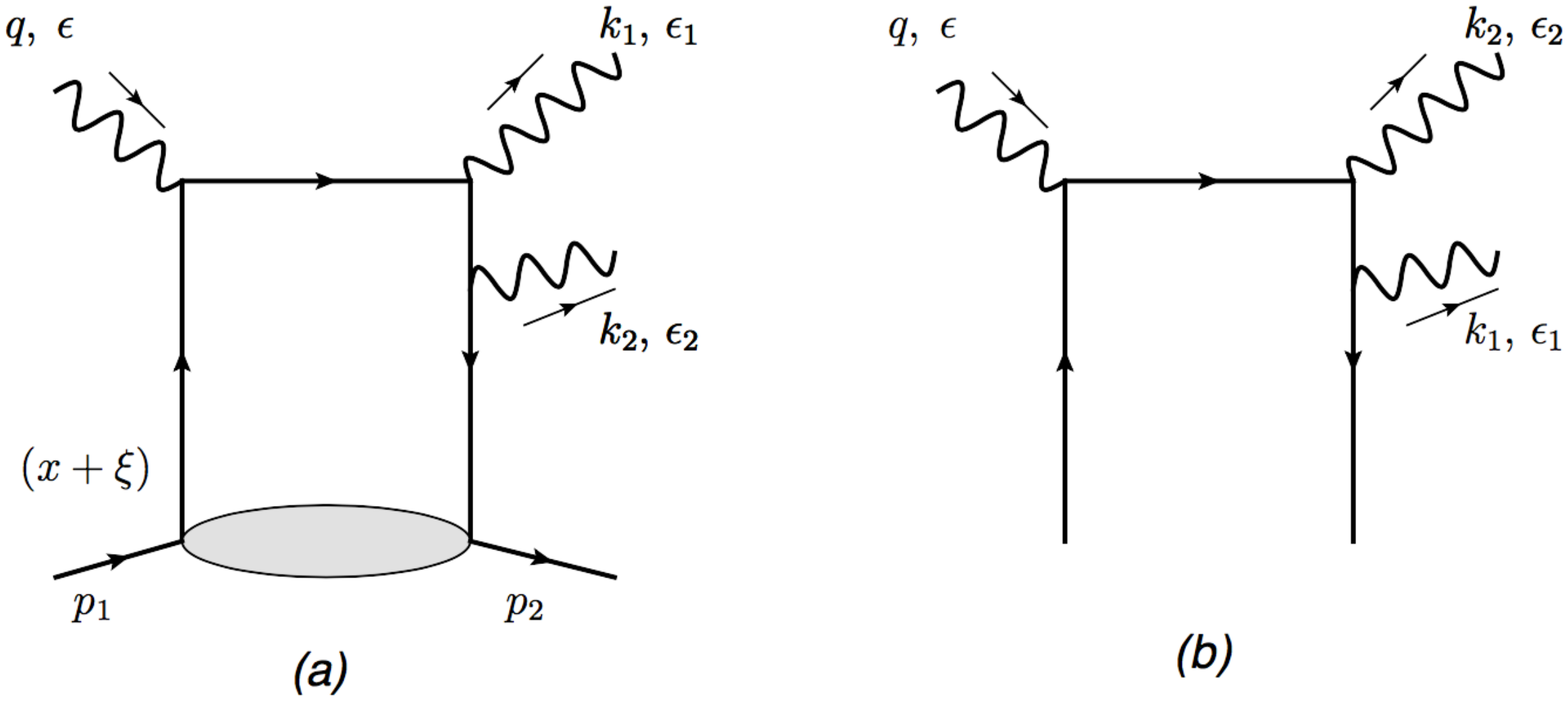}
\includegraphics[width=8cm]{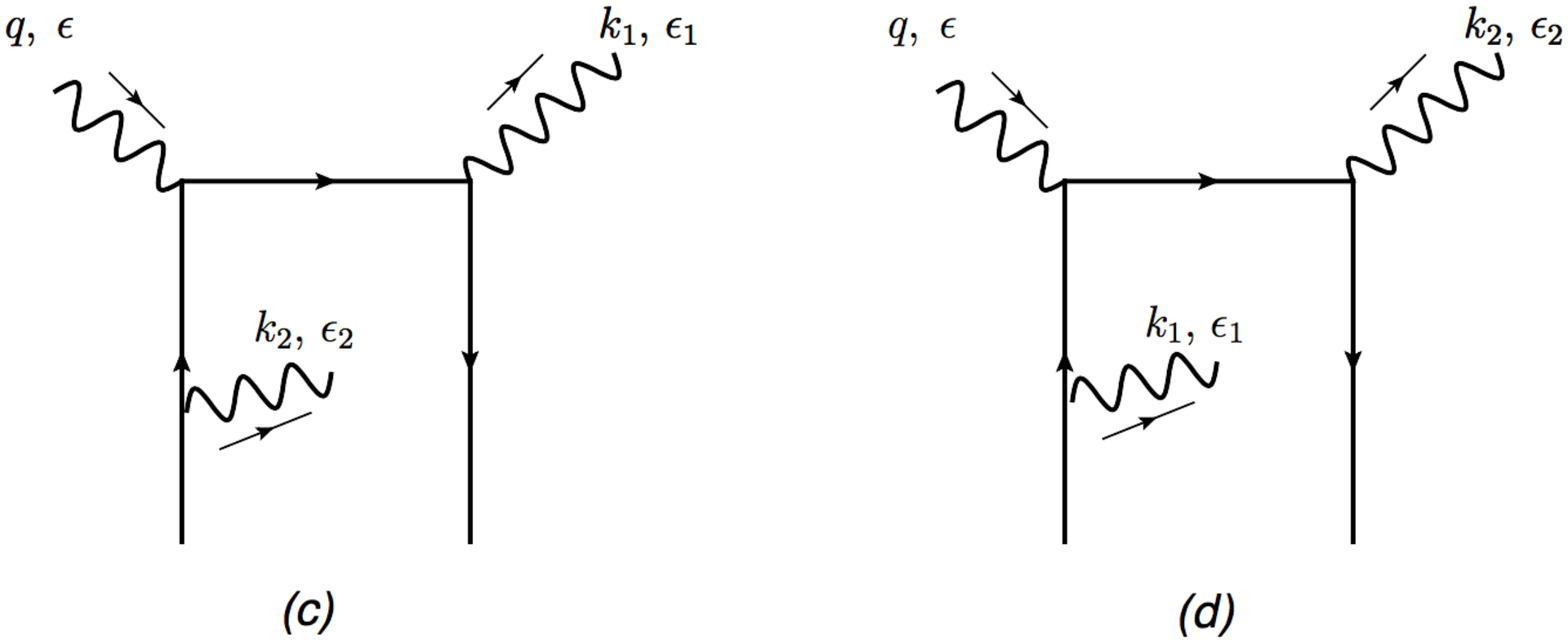}

\center\vspace{-3cm}\includegraphics[width=8cm]{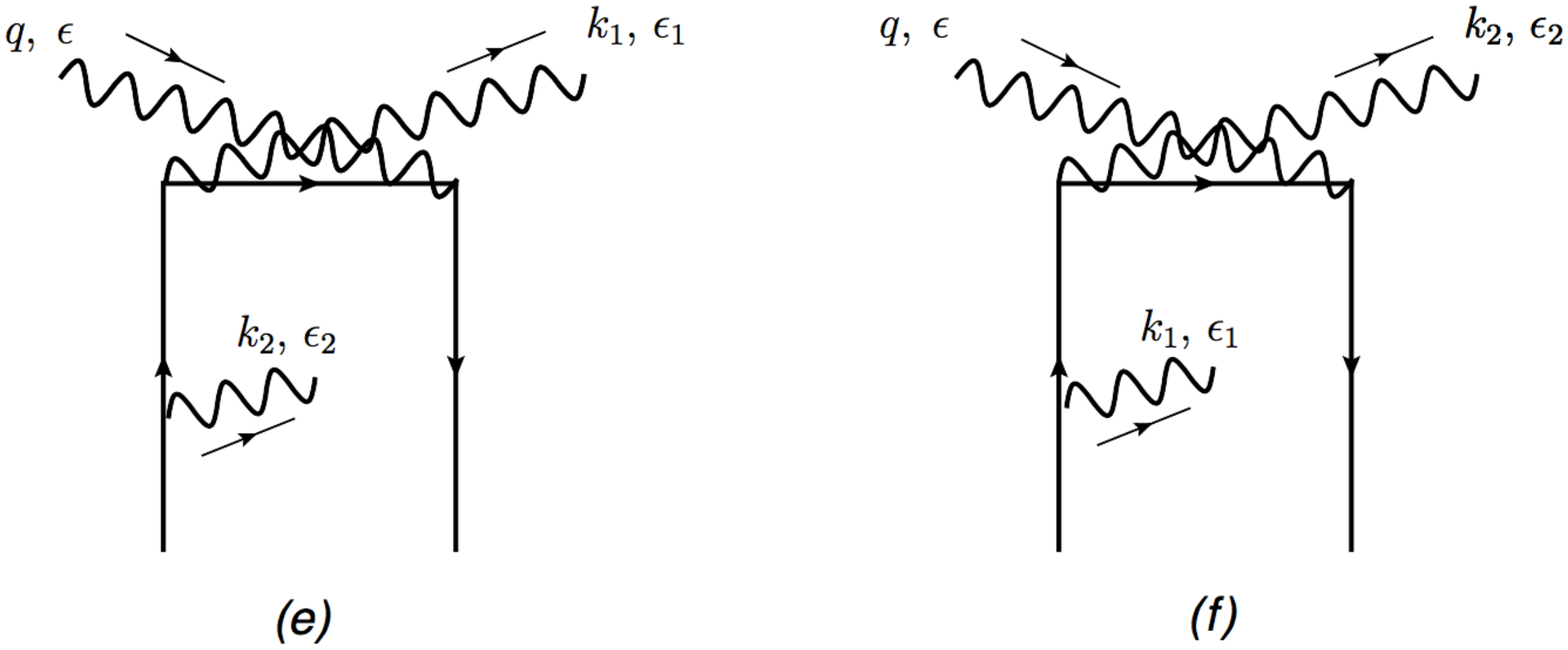}
\vspace{-1cm}
\caption{Feynman diagrams contributing to the coefficient functions of the process $\gamma N \to \gamma \gamma N'$}
\label{feyndiageu3g}
\end{figure} 
Factorization allows to write the scattering amplitude as
\begin{eqnarray}
&\mathcal{T} & =\frac{1}{4P^+}
\int_{-1}^1 dx\sum_q\left[
CF^V_q(x,\xi)\left(
 H^q(x,\xi)\bar{U}(p_2, s_2)\not{n} U(p_1,s_1)+
 E^q(x,\xi)\bar{U}(p_2, s_2)\frac{i\sigma^{\mu\nu}\Delta_\nu n_\mu}{2M}U(p_1, s_1)
\right)
\right.\nonumber \\
&&\left. 
+~~~~~CF^A_q(x,\xi)\left(
\tilde{H}^q(x,\xi)\bar{U}(p_2, s_2)\not{n}\gamma^5 U(p_1, s_1)+
\tilde{E}^q(x,\xi)\bar{U}(p_2, s_2)\frac{i\gamma_5(\Delta\cdot n)}{2M}U(p_1, s_1)
\right)
\right]\, ,
\end{eqnarray}
with the coefficient functions calculated from the diagrams of Fig. 1 as
\begin{eqnarray}
iCF^V_q&=& Tr[i\mathcal{M}\not p]=
-i e_q^3\left[A^V\left(\frac{1}{D_1(x)D_2(x)}+\frac{1}{D_1(-x)D_2(-x)}\right)+\right.\\
&&B^V\left(\frac{1}{D_1(x)D_3(x)}+\frac{1}{D_1(-x)D_3(-x)}\right)+
\left.C^V\left(\frac{1}{D_2(x)D_3(-x)}+\frac{1}{D_2(-x)D_3(x)}\right)\right] \,,
\nonumber\\
iCF^A_q&=& Tr[i\mathcal{M}\gamma^5\not p]=
-i e_q^3\left[A^A\left(\frac{1}{D_1(x)D_2(x)}-\frac{1}{D_1(-x)D_2(-x)}\right)+\right.
\left.
B^A\left(\frac{1}{D_1(x)D_3(x)}-\frac{1}{D_1(-x)D_3(-x)}\right)
\right]\,,
\nonumber
\end{eqnarray}
where ${\mathcal M}\not p$ and ${\mathcal M}\gamma^5\not p$ are contributions of the hard part of scattering amplitude projected on vector and axial vector   Fierz structures, 
 $e_q=Q_q |e|$  and the denominators read
 \begin{eqnarray}
&&D_1(x)=s(x+\xi+i\varepsilon) ~~~,~~~
D_2(x)=s\alpha_2(x-\xi+i\varepsilon)~~~,~~~
D_3(x)=s\alpha_1(x-\xi+i\varepsilon)\,.\nonumber 
\end{eqnarray}
The tensorial structure can be written for the vector part as
\begin{eqnarray}
A^V=2s(V_{k_1}-V_p+\frac{1+\alpha_2}{\alpha_1}V_{k2})~,~
B^V=2s(-V_{k2}+V_p-\frac{1+\alpha_1}{\alpha_2}V_{k_1})~,~
C^V=2s((\alpha_2-\alpha_1)V_p+V_{k_2}-V_{k1})\,,\nonumber
\end{eqnarray}
with 
\begin{eqnarray}
V_{k_1}=(\epsilon_\bot(q)\cdot\epsilon^*_\bot(k_1))
(p_\bot\cdot\epsilon^*_\bot(k_2))~,~
V_{k_2}=(\epsilon_\bot(q)\cdot\epsilon^*_\bot(k_2))
(p_\bot\cdot\epsilon^*_\bot(k_1))~,~
V_{p}=(\epsilon^*_\bot(k_1)\cdot\epsilon^*_\bot(k_2))
(p_\bot\cdot\epsilon_\bot(q))\,,\nonumber
\end{eqnarray}
while the axial part reads
\begin{eqnarray}
A^A=4i\left(A_{k_1}+\frac{1+\alpha_2}{\alpha_1}A_{k_2}-A_p\right)~,~ B^A=4i\left(-\frac{1+\alpha_1}{\alpha_2}A_{k_1}-A_{k_2}+A_p\right) \,,\nonumber
\end{eqnarray}
with
\begin{eqnarray}
A_{k_1}=p_\bot\cdot\epsilon^*_\bot(k_2)\epsilon^{pn\epsilon_\bot(q)\epsilon_\bot^*(k_1)}~,~
A_{k_2}=p_\bot\cdot\epsilon^*_\bot(k_1)\epsilon^{pn\epsilon_\bot(q)\epsilon_\bot^*(k_2)}~,~
A_{p}=\epsilon^*_\bot(k_1)\cdot\epsilon^*_\bot(k_2)\epsilon^{pn\epsilon_\bot(q)p_\bot}\,.\nonumber
\end{eqnarray}

The scattering amplitude is written in terms of generalized Compton form factors $\mathcal{H}^q(\xi)$, 
$\mathcal{E}^q(\xi)$, $\tilde {\mathcal{H}}^q(\xi)$ and 
$\tilde {\mathcal{E}}^q(\xi)$ 
as
\begin{eqnarray}
&\mathcal{T}=&
\frac{1}{2s}
\sum_q
\left[
\left(
\mathcal{H}^q(\xi)\bar{U}(p_2)\not{n} U(p_1)+
 \mathcal{E}^q(\xi)\bar{U}(p_2)\frac{i\sigma^{\mu\nu}\Delta_\nu n_\mu}{2M}U(p_1)
\right)+\right. \nonumber\\
&&
\left.
\left(
\tilde{\mathcal{H}}^q(\xi)\bar{U}(p_2)\not{n}\gamma^5 U(p_1)+
\tilde{\mathcal{E}}^q(\xi)\bar{U}(p_2)\frac{i\gamma_5(\Delta\cdot n)}{2M}U(p_1)
\right)
\right]\,,
\end{eqnarray}

where

\begin{eqnarray}
\mathcal{H}^q(\xi)&=&\int^1_{-1}dx CF^V_q(x,\xi)H^q(x,\xi)=
(-e_q^3)\left[A^V\mathcal{H}^q_{A^V}(\xi)+B^V\mathcal{H}^q_{B^V}(\xi)+C^V\mathcal{H}^q_{C^V}(\xi)\right]\, \nonumber
\\
& = &(-e_q^3) (\alpha_1A^V+\alpha_2B^V) \frac{i\pi}{\xi s^2\alpha_1\alpha_2}(H^q(\xi,\xi)+H^q(-\xi,\xi))\, ,\\
\tilde{\mathcal{H}}^q(\xi)&=&\int^1_{-1}dx CF^A_q(x,\xi)\tilde{H}^q(x,\xi)=
(-e_q^3)\left[A^A\tilde{\mathcal{H}}^q_{A^A}(\xi)+B^A\tilde{\mathcal{H}}^q_{B^A}(\xi)\right] \, \nonumber \\
&=& (-e_q^3) (\alpha_1A^A+\alpha_2B^A)\frac{-i\pi}{\xi s^2\alpha_1\alpha_2}(\tilde H^q(\xi,\xi)-\tilde H^q(-\xi,\xi))\,,
\end{eqnarray}
and similar equations for $\mathcal{E}^q(\xi)$ and $\tilde\mathcal{E}^q(\xi)$.

\section{Cross sections and asymmetries}
The peculiar analytic structure of the coefficient function thus leads to a very interesting and quite unique fact : the leading order cross section of this process is proportional to the (square of the) GPDs at the cross over line $x=\pm \xi$. Contrarily to DVCS or TCS, one does not need to convolute the GPDs with a function. This feature is however not going to survive a NLO analysis which we plan to perform soon.

\begin{figure}[h!]
\center \includegraphics[width=0.4\columnwidth]{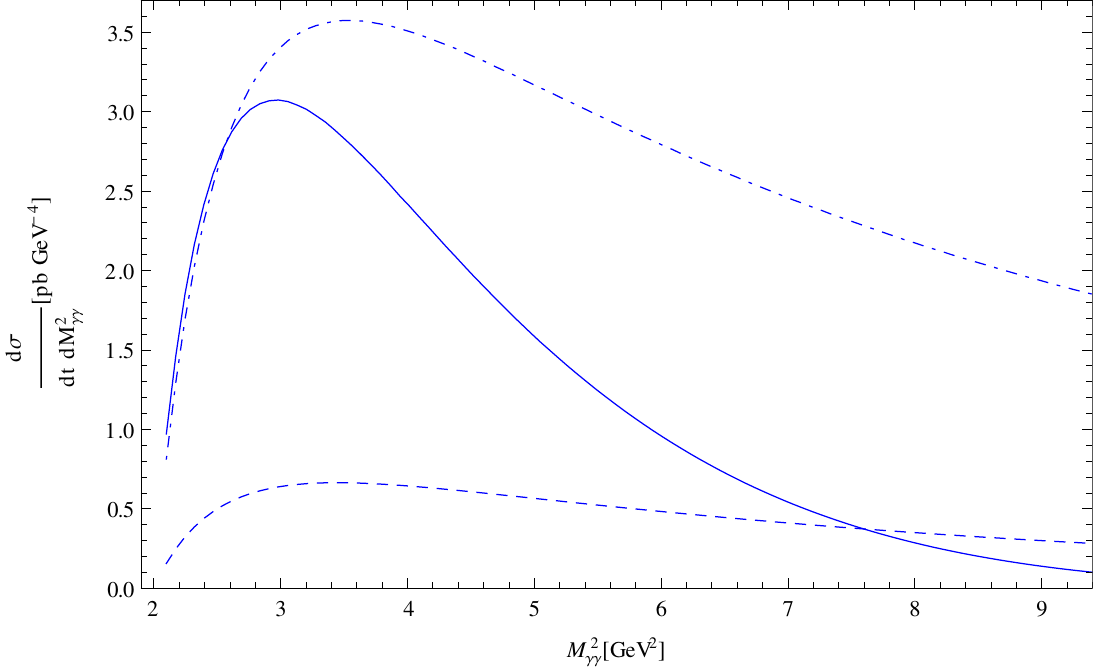}~~~~~
\includegraphics[width=0.4\columnwidth]{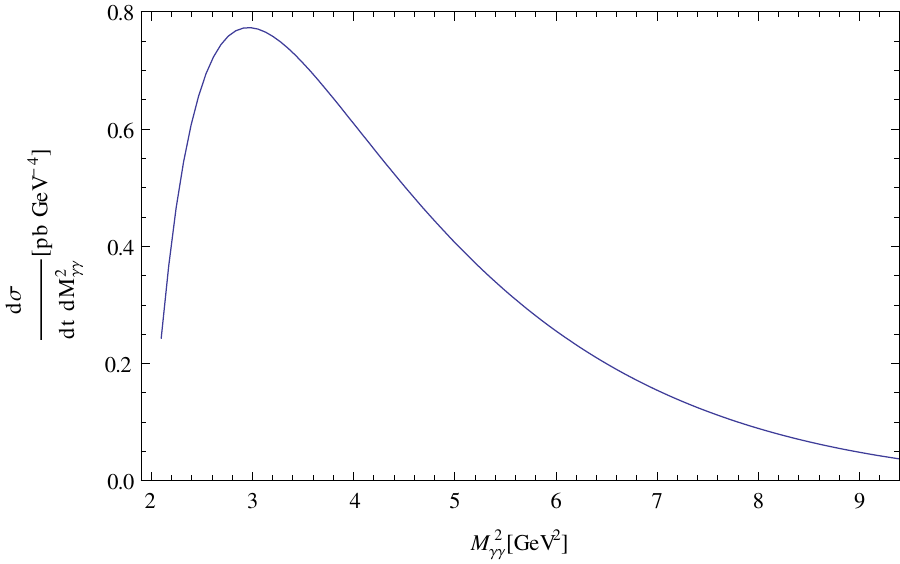}
\caption{The $M_{\gamma\gamma}^2$ dependence of the unpolarized differential cross section $\frac{d\sigma}{dM_{\gamma\gamma}^2dt} $ on a proton(left panel) and on a neutron(right panel) at $t=t_{min}$ and $S_{\gamma N} = 20$ GeV$^2$ (full curves), $S_{\gamma N} = 100$ GeV$^2$  (dashed curve) and $S_{\gamma N} = 10^6$ GeV$^2$  (dash-dotted curve, multiplied by $10^5$).}
   \label{Fig_2}
\end{figure}

 By lack of space, we restrict ourselves to the presentation (Fig. \ref{Fig_2}) of the $M_{\gamma\gamma}^2$ dependence of the unpolarized differential cross section $\frac{d\sigma}{dM_{\gamma\gamma}^2dt} $ on a proton(left panel) and on a neutron(right panel) at $t=t_{min}$ and $S_{\gamma N} = 20$ GeV$^2$ (full curves), $S_{\gamma N} = 100$ GeV$^2$  (dashed curve) and $S_{\gamma N} = 10^6$ GeV$^2$  (dash-dotted curve, multiplied by $10^5$).
    
The conclusion of these cross-section estimates is straightforward. This reaction can be studied at intense photon beam facilities in JLab. The rates are not very large but of comparable order of magnitude as those for the timelike Compton scattering reaction, the feasibility of which has been demonstrated \cite{Boer}. Since there are no contribution from gluons and sea-quarks, one does not get larger cross-sections at higher energies. Contrarily to timelike Compton scattering \cite{PSWUPC}, it thus does not seem attractive to look for this reaction  in ultra peripheral reactions at hadron colliders.

\begin{figure}[h!]
\centering
\includegraphics[height=8cm]{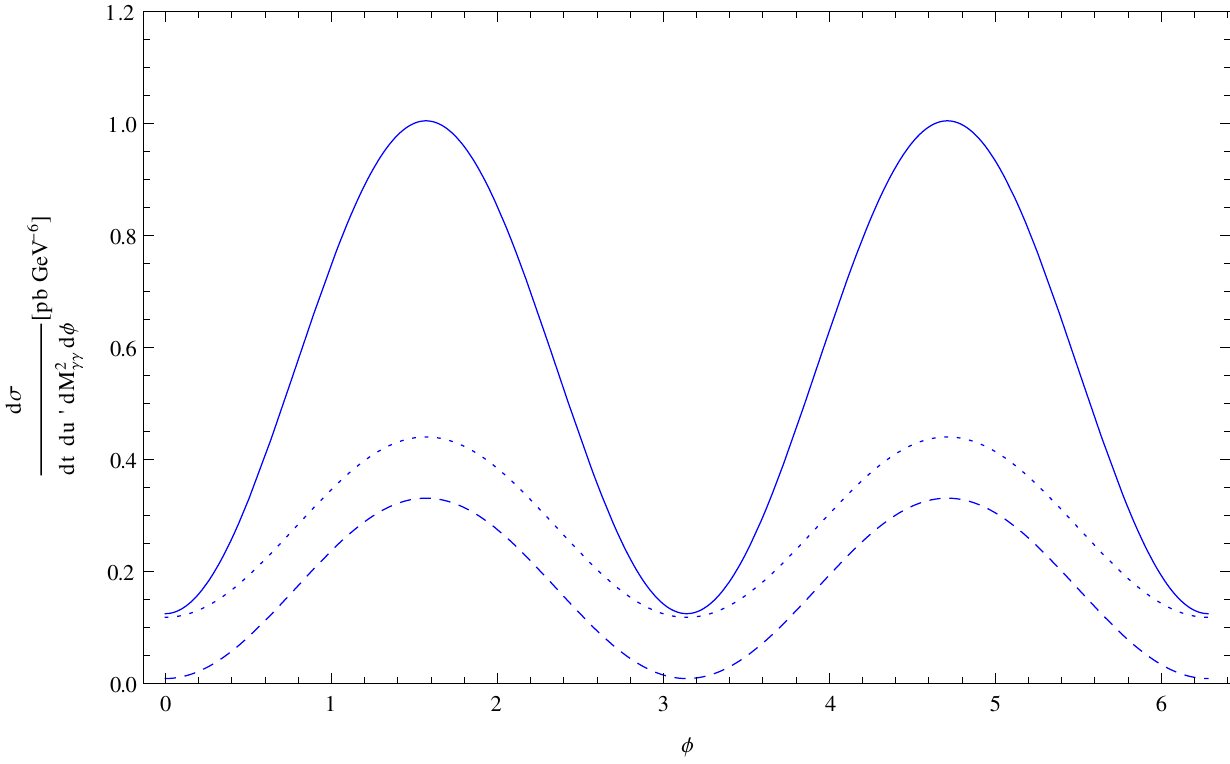}
\caption{The azimuthal dependence of the  differential cross section $\frac{d\sigma}{dM_{\gamma\gamma}^2dt du' d\phi} $ at $t=t_{min}$ and $S_{\gamma N} = 20$ GeV$^2$.  $(M_{\gamma\gamma}^2,u' ) =( 3, -2)$ GeV$^2$  (solid line),  $(M_{\gamma\gamma}^2,u' ) =( 4, -1)$ GeV$^2$  (dotted line) and $(M_{\gamma\gamma}^2,u' ) =( 4, -2)$ GeV$^2$  (dashed line). $\phi$ is the angle between the initial photon polarization and one of the final photon momentum in the transverse plane.}
\label{cs_phi}
\end{figure}

Linearly polarized real photons open the way to large asymmetries, as they do for dilepton photoproduction \cite{Goritschnig:2014eba}. Let us consider the case where the initial photon is polarized along the $x$ axis,  $\epsilon(q) = (0,1,0,0)$, and define the azimuthal angle $\phi$ through 
$$
p_T^\mu = ( 0, ~p_T~ cos \phi, ~p_T ~sin\phi, 0).
$$
The cross section exhibits then  an azimuthal  dependence, and one should calculate 
\begin{equation}
\label{crossecLC}
\frac{d\sigma_{l}}{dM^2_{\gamma\gamma}dtd(-u')d\phi}=\frac{1}{2}\frac{1}{(2\pi)^4 32 S^2_{\gamma N}M^2_{\gamma\gamma}}
\sum_{\lambda_1\lambda_2,s_1,s_2}\frac{|\mathcal{T}|^2}{2}.
\end{equation} 
This is shown on Fig. \ref{cs_phi} for different values of $(M_{\gamma\gamma}^2, u' )$ at $ t=t_{min}$ and $S_{\gamma N} = 20$ GeV$^2$. As straightforwardly anticipated, the cross section shows a modulation of the form $A + B ~cos 2\phi$. It turns out that $B$ is negative leading to  a minimum at $\phi=0$ and a maximum at  $\phi=\pi/2$. In all cases, the linear polarization effects are huge.

This project has received funding from the European Union's Horizon 2020 research and innovation programme under grant agreement No 824093 and from the grant 2017/26/M/ST2/01074 of the National Science Center in Poland.  The  project  is  co-financed  by  the  Polish National Agency for Academic Exchange. L.S. thanks also the French LABEX P2IO, the French GDR QCD and the LPT for support.

\end{document}